\pdfoutput=1 
\documentclass{JINST}
\usepackage{cite}
\usepackage{soul}
\usepackage[normalem]{ulem}
\usepackage{epstopdf}
\usepackage{comment}
\usepackage{multirow}
\usepackage{float}
\title{Study of neutron-induced background and its effect on the search of 0$\nu\beta\beta$ decay in $\rm^{124}Sn$}

\author{N.~Dokania$^{a,b}$, V.~Singh$^{a,b}$, S.~Mathimalar$^{a,b}$, C.~Ghosh$^c$,  
V.~Nanal$^c$\thanks{Corresponding author.}, R.G.~Pillay$^c$, S.~Pal$^c$, K.G.~Bhushan$^d$
and A.~Shrivastava$^e$\\
\llap{$^a$}India-based Neutrino Observatory,\\
 Tata Institute of Fundamental Research, Mumbai 400 005, India \\
\llap{$^b$}Homi Bhabha National Institute,\\
  Anushaktinagar, Mumbai 400 094, India\\
\llap{$^c$}Department of Nuclear and Atomic Physics, Tata Institute of Fundamental Research,\\
Colaba, Mumbai 400 005, India\\ 
 \llap{$^d$} Technical Physics Division, Bhabha Atomic Research Centre,\\ Mumbai 400 085, India\\
  \llap{$^e$} Nuclear Physics Division, Bhabha Atomic Research Centre,\\ Mumbai 400 085, India\\
 
E-mail: \email{nanal@tifr.res.in}}

\abstract{Neutron-induced background has been studied in various components of the TIN.TIN detector, which is under development for the search of Neutrinoless Double Beta Decay in $\rm^{124}Sn$. Fast neutron flux $\sim10^{6}~n~cm^{-2}s^{-1}$ covering a broad energy range ($ \sim0.1$ to $ \sim18$~MeV) was generated using $^{9}Be(p,n)^{9}B$ reaction. In addition, reactions with quasi-monoenergetic neutrons were also studied using $^{7}Li(p,n)^{7}Be$ reaction. Among the different cryogenic support structures studied, Teflon is found to be preferable compared to Torlon as there is no high energy gamma background ($E_\gamma >$ 1 MeV). Contribution of neutron-induced reactions in $\rm ^{nat, 124} $Sn from other Sn isotopes (A = 112 -- 122) in the energy region of interest, namely, around the $Q_{\beta\beta}$ of $\rm^{124}Sn$ ($E \sim$ 2.293 MeV), is also investigated.}


\begin{document}

\section{Introduction}\label{sec:intro}
For rare event studies like Double Beta Decay (DBD) and dark matter searches, the reduction of background is very important for improving the sensitivity of the experiment. Typical half-life for DBD process, $T_{1/2}$ $>$10$^{18}$ years~\cite{barabash, elliott, cremonesi}, is much larger than that of natural radioactivity from U, Th chains and $ ^{40}\rm K $ ($T_{1/2}\sim 10^{8} - 10^{10}$~years). In recent years, ultra-low levels of background $<10^{-2}$~counts/(keV kg yr) have been reported in rare decay event experiments~\cite{gerda, cuore, exo, nemo3, kamland}. The background level in an underground laboratory is mainly limited by trace impurities present in detector materials, which can be minimized but cannot be eliminated completely. Of the different sources of background, namely, $\alpha, \beta, \gamma$ and neutrons, background arising from neutrons is most difficult to suppress and hence crucial to understand. In fact, neutrons are reported to be the limiting source of background for dark matter search experiments since they can produce nuclear recoils via elastic scattering off target nuclei resulting in a signal similar to that of WIMPs (Weakly Interacting Massive Particles)~\cite{kim, cdms, lux}. Neutrons are produced in the spontaneous fission of $\rm^{nat}$U (mainly $^{238}$U),~Th present in the rocks and the surrounding materials. In addition, alpha particles produced from decay of intermediate nuclei in the natural decay chains can react with light nuclei in the rocks to produce neutrons via ($\alpha$, n) reactions~\cite{fiss}. Very high energy neutrons ($E_n\sim$ GeV) are produced by muon-induced interactions in the rocks and materials surrounding the detector. It has been reported that in an underground laboratory, the low energy neutron flux ($E_n <$ 10~MeV) from natural radioactivity is about two to three orders of magnitude higher than that from the muon-induced reactions~\cite{kim, bellini1, fiss, mei}. Although the high energy neutrons are more penetrating, the average neutron energy reduces from 100 -- 200~MeV to $ \sim $ 45~MeV~\cite{lead} as they propagate through layers of shield materials. 
It should be mentioned that active veto systems are employed for the rejection of muon-induced events in the shield and detector/source assembly~\cite{bellini}, while the flux of the low energy neutrons is reduced by suitable shielding (mostly hydrogenous materials) around the detector~\cite{bungau}. Thus, it is important to understand the background arising from low energy neutrons. With low energy neutrons, the inelastic scattering of neutrons (n, n$'\gamma$) and neutron-capture (n, $\gamma$) with the source/detector and the surrounding materials are main sources of gamma background. Moreover, these neutrons after thermalisation in the shield can produce significant background by radiative capture reactions in the detector/source assembly.
In addition, any impurities in these materials could be potential sources of neutron-induced background.
The reaction products formed upon neutron activation can have half-lives ranging from $\sim \rm min$ to $\sim \rm years$. The short-lived activities can be avoided by storing the material for prolonged periods in underground locations but the long-lived activities are highly undesirable.

In India, TIN.TIN detector (The INdia-based TIN detector) comprising cryogenic bolometer array of Tin detector elements is under development for a feasibility study to search for $0\nu\beta\beta$ decay in $\rm^{124}$Sn~\cite{inpc, pramana}. It is essential to understand the gamma background in the region of interest (ROI) near $Q_{\beta\beta}$, which for $\rm^{124}Sn$ is
2292.64$\pm$0.39~keV~\cite{qvalue}. With this motivation, the neutron-induced background ($E_n <$ 20~MeV) in TIN.TIN detector components is investigated. The aim of the neutron activation study ($E_n <$ 20~MeV) is two fold -- the selection of materials suitable for use in and around the cryogenic bolometer and the evaluation of its effect on the gamma background level. The paper is organized as follows: Section 2 describes the experimental details, while data analysis and results are presented in Section 3. Conclusions are discussed in Section 4.

\section{Experimental Details}\label{sec:expt}

The TIN.TIN detector will consist of $\rm^{nat}$Sn or $\rm^{124}Sn$ bolometer mounted in a specially designed low background cryostat. The neutron-induced gamma background from the cryostat housing can be significantly reduced by mounting low activity Pb shield inside the cryostat (similar to CUORE~\cite{cuoredesign}). Hence, only the neutron activation of materials in the close vicinity of the detectors elements is of prime importance.  
For neutron-induced background study the materials chosen are: high purity ETP (Electrolytic Tough Pitch) $\rm^{nat}$Cu~(2N purity) used inside the cryostat; Torlon (4203), Torlon (4301) and Teflon -- cryogenic materials for detector holders; $\rm^{nat}$Pb -- the common shielding material, $\rm^{nat}Sn$~(7N purity) and 97.2$\%$ enriched $\rm^{124}Sn$ (research grade, supplied by M/S Isoflex~\cite{isoflex}). It is known that the Torlon has better tensile strength and lower coefficient of linear thermal expansion  but higher thermal conductivity as compared to the Teflon. In addition to the thermal properties, the radiopurity and neutron-induced background is an important factor for choice of the detector holder material in cryogenic bolometer.
Torlon 4203, 4301 and Teflon samples used were of standard commercial grade. Elemental concentrations were obtained using Time of Flight Secondary Ion Mass Spectrometry (TOF-SIMS).
Since all the materials contain high percentage of $ ^{19} $F, Secondary Ion Mass Spectra were obtained in both positive and negative ion modes to ascertain the total fluorine concentration. Final elemental concentrations were obtained after suitable correction with weighted relative sensitivity factors (RSF) for individual element~\cite{wilson}. Besides C, F and O, the major elements found in Torlon 4203 is Ti (contains TiO$_2$ ~\cite{solvay}) while Fe was found in Torlon 4301, which could be undesirable for low temperature applications.


The neutron activation was performed using proton beam on Be and Li production targets in the neutron irradiation setup at the Pelletron Linac Facility, Mumbai~\cite{sharma}. Irradiation targets were mounted in a forward direction with respect to the proton beam, close to the production target but outside the vacuum chamber. This facilitated the change of irradiation targets without breaking the accelerator vacuum. The setup is located in a well shielded area above the analyzing magnet of the Pelletron, which permits the use of high proton beam current $\sim$ 120~nA on the production target. In the present study, proton beams of energy $E_p$ = 10, 12 and 20~MeV on a Be target (5~mm thick) were used to obtain neutrons of a broad energy range with reaction $\rm^{9}Be(p,n)\rm^{9}B$ (Q = --1.850 MeV)~\cite{kamada}. Beam energies were chosen to cover the energy range of neutron spectra originating from fission and ($\alpha$, n) reactions in the rocks ~\cite{fiss}. The energy dependence of the cross-sections of the possible reaction channels in different targets was also taken into consideration. 
 In addition, nearly mono-energetic neutrons were produced with the $\rm^{7}Li(p,n)\rm^{7}Be$ (Q = --1.644 MeV) reaction by bombarding a 0.15~mm thick natural Lithium target (wrapped in a $\sim2~\mu$ thick Ta foil) with proton beam of energy 12~MeV. Contribution from the $\rm^6Li$ (natural abundance 7.59$\%$) in the natural lithium target is expected to be negligible. At $E_p$ = 12~MeV, due to the contributions from the excited states of $\rm^{7}Be$, quasi-monoenergetic neutrons are produced~\cite{poppe, simakov}. The flux obtained in case of the Li target was smaller than that in the case of Be by a factor of $\sim10$. However, the better definition of neutron energy was useful for identification of some of the reaction channels. It should be mentioned that neutron flux could not be measured accurately in the setup and hence $\rm^{nat}$Fe target ($\sim$ 5 -- 6~mg/cm$^{2}$) was used to estimate neutron flux with the $\rm^{56}Fe(n,p)\rm^{56}Mn$ reaction.
Multiple irradiation targets (upto five) were stacked in a 3~cm long target holder (Aluminum or Teflon) using Teflon spacers for an efficient utilization of beam time. 
In this geometry, the solid angles subtended by the neutron beam at the first and last target were $\sim$ 0.25 sr and $\sim$ 0.04 sr, respectively. Thickness of irradiated targets varied from 1.8 mg/cm$^{2}$ to 0.29 g/cm$^{2}$. Both short (2 -- 3 h) as well as long (10 -- 35 h) duration irradiation were carried out to look for short-lived and long-lived products. 
 In case of long irradiation experiments, the access to target area was restricted due to the radiation safety limits and targets could be taken out for measurements only after sufficient cooling time ($ \sim $ 20~min to $ \sim $ 1~h). Hence, some of the short-lived activities could not be observed.
 
The irradiated targets were counted offline for the detection of characteristic $\gamma$-rays of reaction products resulting from neutron activation. Three counting setups with efficiency calibrated HPGe detectors were used. One setup consisted of a low background detector of relative efficiency (R.E.) $\sim$ 70$\%$ with a 10~cm low activity Pb ($^{210}$Pb $<$ 0.3~Bq/kg) and 5~cm low activity Cu shield \cite{neha}. The other two HPGe detectors of R.E. $\sim$ 30$\%$ (D1 and D2) were shielded with 5~cm thick normal Pb rings. D1 and D2 were mostly used for identification of gamma-rays and half-life measurements.
Targets were mounted in a close geometry in these counting setups to search for low levels of activity and coincident summing effects had to be taken into account. Data were recorded with a commercial FPGA (Field-Programmable Gate Array) based 100~MS/s digitizer (CAEN-N6724)~\cite{anoop} and analyzed using LAMPS \cite{lamps}. It should be mentioned that all targets were studied in the low background setup prior to irradiation and did not show any radioactivity above the background level. 

\label{sec:flux}
\subsection {Estimation of neutron flux}
As mentioned earlier, the neutron flux is estimated from the yield of 846.7~keV $\gamma$-ray, produced via $\rm^{56}Fe(n,p)\rm^{56}Mn$ reaction. Since the neutron spectra produced from the $\rm^9Be(p,n)\rm^9B$ reaction is continuous, energy integrated neutron flux has been estimated in the energy range of $E_n$ $ \sim $ 0.1 MeV to $E_{max}$, where $E_{max}= E_p -Q_{th}$ with $Q_{th}=2.057\rm~MeV$. 

The number of Mn atoms (${N_{Mn}}$) produced by irradiation of a Fe target with a constant neutron flux $\phi _{n} $ ($ n~cm^{-2}~s^{-1} $) for time $ t_{irr} $ is given by,

\begin{equation}
\label{eqn1}
N_{Mn} =\frac{{N_{Fe}}~(1-e^{-\lambda t_{irr}})~\sum_{E_n}\!{{\sigma_{c}(E_n)\phi_n(E_n)}dE_n}}{\lambda}\\
\end{equation}
where $ N_{Fe}$ is number of Fe target atoms, $ \lambda $ is the decay constant of~$ ^{56} $Mn and ${\sigma_{c}(E_n)}$ is the $(n,p)$ cross-section of $\rm^{56}Fe(n,p)\rm^{56}Mn$ reaction at the neutron energy $ E_{n} $. The factor $((1-e^{-\lambda t_{irr}})/\lambda)$ arises from decay during irradiation. 

The $N_{Mn} $ can be obtained from the measured photo peak area (${N_{\gamma}}$) of 846.7~keV $\gamma$-ray as, 

\begin{equation}
\label{eqn2}
{N_{Mn}} =\frac{N_{\gamma}} {{e^{-\lambda t_c}~(1-e^{-\lambda t)}~I_{\gamma}~\epsilon_{\gamma}}}\\
\end{equation}
where $ t_{c}$ is the time elapsed between the end of neutron irradiation and start of the counting (cool-down time), $t$ is the counting period, $I_\gamma$ is the branching ratio and $\epsilon_\gamma$ is the photo peak detection efficiency of $E_{\gamma}$ (846.7~keV) for a finite size source in close geometry, computed from Geant4-based Monte Carlo simulations~\cite{neha}. 

Since the distribution of neutrons produced from the $\rm^9Be(p,n)\rm^9B$ reaction is continuous, energy integrated neutron flux can be estimated as,
\begin{equation}
\label{eqn3}
{<\phi_n>} =\frac{\sum_{E_n}\!{{\sigma_{c}(E_n)}\phi_n(E_n)\, dE_n}}{\sum_{E_n}\!{{\sigma_{c}(E_n)}\, dE_n}}
\end{equation}

The numerator in eq.~3.3 is extracted from eq.~3.1 while the denominator is obtained using ENDF/B-VII library~\cite{nndc}. 
In case of Li target, since the emitted neutrons are nearly monoenergetic, the measured value of neutron capture cross section in the same setup, $\sigma_c=$ 65.88 (4.54) barn, at an average neutron energy $E_n$ = 9.85~MeV corresponding to $E_p$ = 12~MeV is used~\cite{iron}.

Table~\ref{flux} shows the extracted neutron flux at a distance $d\sim 5\rm~cm$ from the production target (Be/Li) for different proton energies together with maximum energy of the neutrons $ E_{max}$, average energy of the neutrons $<E_{n}>$ and the average proton beam current $ <I> $. The $ <E_n> $ for $p + \rm^{9} $Be reaction is calculated as,

\begin{equation}
\label{eqn4}
{<E_{n}>} =\frac{\sum_{i=0}^{3}\sum_{E_n}\!{{\sigma_{(p,n_i)}(E_n)}E_n\, dE_n}}{\sum_{i=0}^{3}\sum_{E_n}\!{{\sigma_{(p,n_i)}(E_n)}\, dE_n}}
\end{equation}
where the summation runs over $E_n$ from $ \sim0.1$~MeV to $E_{max}$ and $ \sigma_{(p,n_i)} (E_n)$ corresponds to\\ 
$\rm^{9}Be(p,n_i)\rm^{9}B $ cross-section at $E_n$ for the $i$th channel of neutron production~\cite{nndc}. Only $(p,n_{0}) $, $(p,n_{1}) $, $(p,n_{2}) $ and $(p,n_{3})$ channels are considered and others with total cross-sections $<6\% $ of ($p,n_{0} $) are neglected.
The uncertainty in the neutron flux includes the error in $\epsilon_{\gamma}$, statistical and fitting errors in the photo peak area of 846.7~keV $ \gamma $-ray ($N _{\gamma} $) and error in the coincident summing correction factor for $ ^{56} $Mn. It should be noted that the neutron flux at $ E_{p} =10~\rm MeV$ could not be measured since the activity of 846.7~keV $ \gamma $-ray was not observed due to relatively lower yield.

\begin{table}[H]
\centering
\caption{\label{flux} Estimated energy integrated neutron flux from $\rm^{56}Fe(n,p)\rm^{56}Mn$ reaction for 12 and 20 MeV proton energies (at $d\sim$ 5 cm). }
\begin{tabular}{|c |c |c| c|c|c|}
\hline
Production Target&$E _{p} $  &$E_{max}$  &$<E_{n}>$ &$ \phi_n $&$<I>$\\ 
 & (MeV)&(MeV)& (MeV)&($ n~cm^{-2}s^{-1}$)& (nA)\\ \hline

\multirow{2}{*}{$\rm^{9}Be$}	&	12	&	9.9 & 3.9 & 2.3(0.2)$\times$10$ ^5 $ & 133\\
			&	20	& 17.9  & 5.6 &	9.9(0.7)$\times$10$ ^5 $ & 148\\ \hline

$\rm^{nat}Li$~\cite{naik} &	12	&	10.1  & 9.85 & 1.3(0.2)$\times$10$ ^5 $ & 112\\ \hline
\end{tabular}
\end{table}

\section{Data Analysis and Results}\label{sec:results}

Table~\ref{listBe} lists the details of the products formed in different samples together with their half-lives and the expected most intense $\gamma$-rays. The last two columns of the table~\ref{listBe} show the minimum neutron energy $E_n$ at which the cross-section for the respective neutron-induced reaction channel is $\ge\mu$b. In most of the cases, the half-life ($T_{1/2}$) of the reaction products were measured and were found to agree within 20$\%$ of the reference values~\cite{nndc}. As the expected energy resolution of the Tin bolometer is 0.2--0.5\% (full width at half maximum) at $Q _{\beta\beta}$, the ROI for background estimation is taken as 2292.6 $\pm$ 25~keV (i.e., $Q _{\beta\beta} \pm 5\sigma$).
The gamma-rays with energies within this ROI as well as with E $\ge Q _{\beta\beta} $ are potential sources of background and are highlighted in bold text in table~\ref{listBe}. It should be mentioned that many of these reaction products decay by $ \beta^-$ emission and if the $ Q_{\beta}\ge$ than the $ Q_{\beta\beta}(^{124} $Sn), electrons or bremsstrahlung resulting from these electrons can contribute to the background in ROI. In particular, the (n,$ \gamma $) reaction on $ ^{124} $Sn leads to $ ^{125} $Sn which $ \beta^-$ decays with a $ Q_{\beta} $ (2357~keV) value close to the $ Q_{\beta\beta} $ of $ ^{124} $Sn. Due to short range of electrons, contribution to the background in the detector arising due to $\beta$-decays in the shield and support materials will be mainly from the surface events. This together with $\beta$-decays within the detector will affect the background, which is not considered in the present work. 

 \begin{table}[H]
\centering
\caption{\label{listBe}Neutron-induced reaction products, $T_{1/2}$ and the expected most intense $\gamma$-rays in the irradiated samples. The minimum neutron energy $E_n$ at which corresponding $\sigma$ is $\ge\mu$b is also listed~\cite{nndc}. }
\smallskip
\begin{tabular}{|c | c |c| c|c|c|}
\hline
 Sample&Reaction & T$_{1/2}$ & $E_{\gamma}$ & $E_{n}$ & $\sigma$\\ 
 &channel &  &  (keV) & (MeV) & (barn)\\ \hline\hline

\multirow{6}{*}{Torlon 4203}&$\rm^{nat} Ti(n,X)^{47}Sc$& 3.3492 d& 159.4 & & \\ \cline{2-6}
& \multirow{2}{*}{$\rm^{nat} Ti(n,X)^{48}Sc$}& \multirow{2}{*} {43.67 h}& 175.4, 983.5 & & \\
& & & 1037.5, 1312.1 & &\\ \cline{2-6}
&$\rm^{nat}Ti(n,X)^{46}Sc$& 83.79 d& 889.3, 1120.5&
 & 
 \\ \cline{2-6}
&$\rm^{27}Al(n,\alpha)^{24}Na$& 14.997 h& 1368.6, \textbf{2754.0} & 4.6& 1.4$\times10^{-6}$\\ \cline{2-6}
&$\rm^{27}Al(n,p)^{27}Mg$& 9.458 min& 843.8, 1014.5& 2.5& 1.9$\times10^{-5}$\\ \hline\hline
 
\multirow{3}{*}{Torlon 4301} & \multirow{2}{*}{$\rm^{56}Fe(n,p)^{56}Mn$}& \multirow{2}{*}{ 2.5789 h}& 846.8, 1810.7  &\multirow{2}{*}{4} &\multirow{2}{*}{ 6.0$\times10^{-6}$}\\

& & & 2113.1 & & \\\cline{2-6}
&$\rm^{27}Al(n,\alpha)^{24}Na$& 14.997 h& 1368.6, \textbf{2754.0} &4.6 & 1.4$\times10^{-6}$ \\ \hline\hline

Teflon &
$\rm^{19}F(n,2n)^{18}F$& 109.77 min& 511 & 11.5&1.5$\times10^{-3}$ \\\hline\hline

\multirow{5}{*}{$\rm ^{nat} $Pb}&
$\rm^{204}Pb(n,2n)^{203}Pb$& 51.92 h& 279.2, 401.3, 680.5 & 8.5&2.1$\times10^{-3}$ \\\cline{2-6}

&\multirow{2}{*}{$\rm^{204}Pb(n,n')^{204m}Pb$}& \multirow{2}{*}{66.93 min}& 374.8, 899.2, &1.0 &2.4$\times10^{-1}$ \\
 && & 911.7, 1274 & & \\\cline{2-6}

&$\rm^{121}Sb(n,\gamma)^{122}Sb$& 2.7238 d& 564.2, 692.7 &0.1 &2.1$\times10^{-1}$ \\\cline{2-6}

& \multirow{2}{*}{$\rm^{123}Sb(n,\gamma)^{124}Sb$}&  \multirow{2}{*}{60.20 d}& 602.7, 1690.9& \multirow{2}{*}{0.1} &\multirow{2}{*}{1.9$\times10^{-1}$ }\\
& &&2090.9, 2182.6, \textbf {2294.0} & &\\ \hline\hline

\multirow{7}{*}{$\rm ^{nat} $Cu}&
$\rm^{63}Cu(n,\gamma)^{64}Cu$& 12.701 h&511, 1345.8&0.055&2.5$\times10^{-2}$  \\ \cline{2-6}

&$\rm^{63}Cu(n,\alpha)^{60}Co$& 1925.28~d&1173.2, 1332.5&2.5&1.1$\times10^{-2}$  \\ \cline{2-6}

&$\rm^{65}Cu(n,\gamma)^{66}Cu$& 5.120 min&1039.2&0.06&1.1$\times10^{-2}$  \\\cline{2-6} 

& \multirow{2}{*}{$\rm^{65}Cu(n,\alpha)^{62m}Co$}&  \multirow{2}{*}{13.91 min}&1163.5, 1172.9, {2003.7} & \multirow{2}{*}{5}& \multirow{2}{*}{3.9$\times10^{-6}$}  \\
&& & 2104.9, \textbf{2301.9, 2882.3 }&& \\ \cline{2-6}

&$\rm^{65}Cu(n,p)^{65}Ni$&  \multirow{2}{*}{2.5175 h}&1115.5, 1481.8, &2.5&1.0$\times10^{-6}$ \\ 

&$\rm^{64}Ni(n,\gamma)^{65}Ni$& &1623.4, 1724.9&0.553&6.5$\times10^{-3}$ \\ \hline\hline

\multirow{15}{*}{$\rm ^{nat, 124} $Sn}&

$\rm^{112}Sn(n,np)^{111}In$& 2.8047 d&171.3, 245.4 & 12 & 4.9$\times10^{-6}$\\\cline{2-6}

&$\rm^{116}Sn(n,np)^{115m}In$&\multirow{3}{*}{ 4.486 h}&\multirow{3}{*}{336.2}&14.5& 7.6$\times10^{-5}$\\

&$\rm^{115}Sn(n,p)^{115m}In$&&&5& 1.9$\times10^{-4}$\\
&$\rm^{115}In(n,n')^{115m}In$&&&0.5& 6.4$\times10^{-3}$\\\cline{2-6} 

 &\multirow{2}{*}{$\rm^{116}Sn(n,p)^{116m}In$}& \multirow{2}{*}{ 54.29 min}&416.9, 818.7, 1097.3& \multirow{2}{*}{8} &\multirow{2}{*}{ 1.5$\times10^{-4}$}\\
& & & 1293.6, {2112.3} &  & \\\cline{2-6}
 
 &$\rm^{117}Sn(n,n')^{117m}Sn$&\multirow{3}{*}{ 13.76 d}&\multirow{3}{*}{ 156.0, 158.6}&0.2&2.9$\times10^{-1}$ \\

&  $\rm^{116}Sn(n,\gamma)^{117m}Sn$& &&0.1&5.5$\times10^{-2}$ \\

& $\rm^{118}Sn(n,2n)^{117m}Sn$& &&9.9&7.2$\times10^{-2}$ \\\cline{2-6} 
 
 &$\rm^{124}Sn(n,2n)^{123m}Sn$&\multirow{2}{*}{40.06 min}&\multirow{2}{*}{160.3}&9&1.6$\times10^{-1}$ \\
 
 & $\rm^{122}Sn(n,\gamma)^{123m}Sn$& &&0.3&1.2$\times10^{-2}$ \\\cline{2-6}
  
&$\rm^{124}Sn(n,2n)^{123}Sn$&\multirow{2}{*}{129.2~d}&\multirow{2}{*}{1088.6}&9&1.6$\times10^{-1}$ \\
&$\rm^{122}Sn(n,\gamma)^{123}Sn$&&&0.3&1.2$\times10^{-2}$ \\ \cline{2-6}

& $\rm^{124}Sn(n,\gamma)^{125m}Sn$ & 9.52 min&331.9&\multirow{2}{*}{0.315}& \multirow{2}{*}{6.8$\times10^{-3}$ }  \\

&$\rm^{124}Sn(n,\gamma)^{125}Sn$& 9.64 d&822.5, 1067.1, 1089.2&&\\\hline
\end{tabular}
\end{table}

\subsection{Neutron-induced activity from Torlon and Teflon}

\begin{figure}[H]
\includegraphics[scale=0.6]{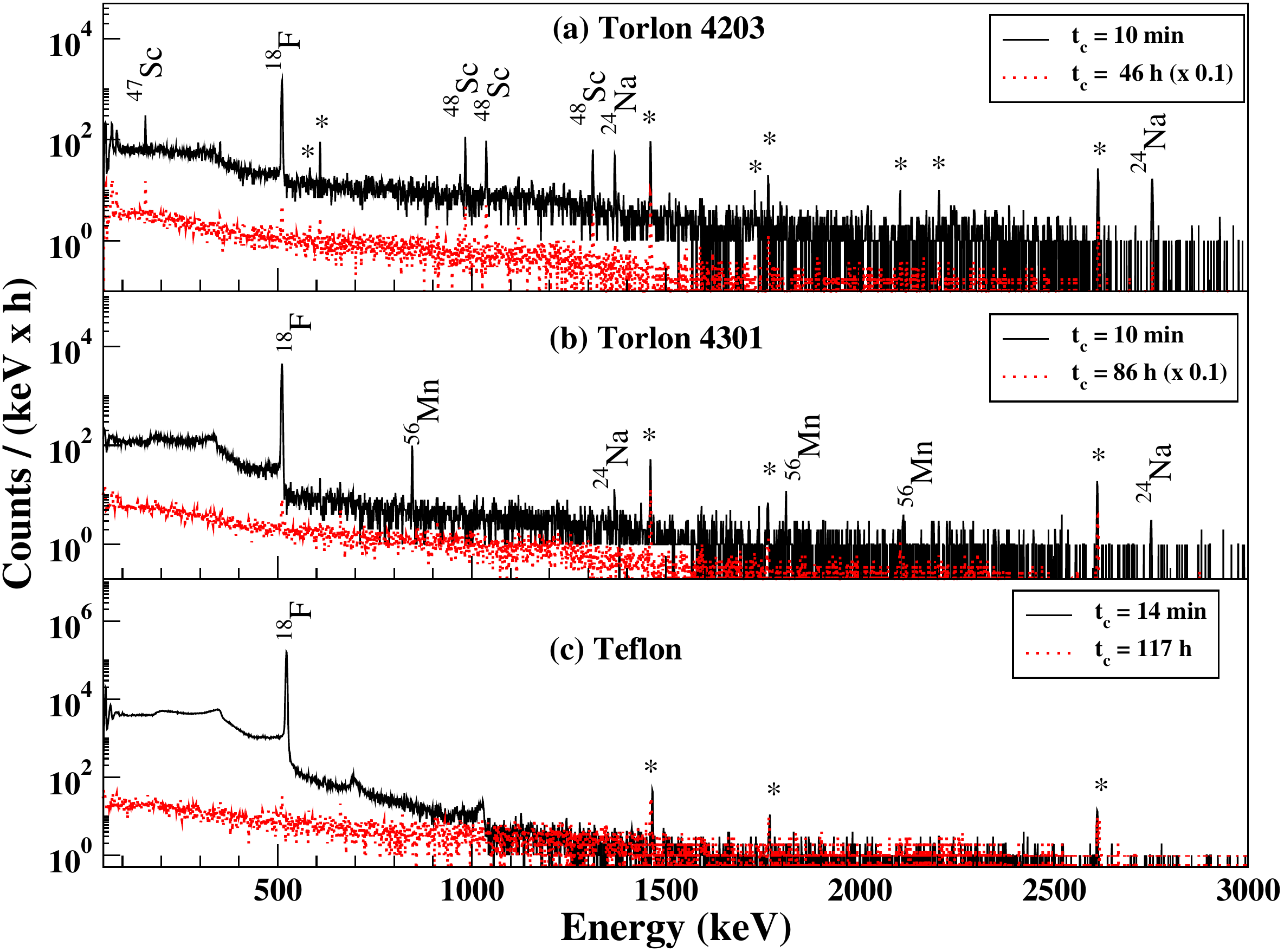} 
\centering
\caption{\label{fig1}(Color online) $\gamma$-ray spectra of the  neutron irradiated (a) Torlon 4203 with $E_p$ = 20~MeV, $t_{irr}$ = 3~h, $N_ n\sim 1.31(9)\times10^{10}$, (b) Torlon 4301 with $E_p$ = 20~MeV, $t_{irr}$ = 3 h, $ N_n\sim 1.38(9)\times10^{10}$ and (c) Teflon with $E_p$ = 20~MeV, $t_{irr}$ = 2~h, $ N_n\sim 0.82(6)\times10^{10}$ for different $t_c$ -- time elapsed since the end of irradiation. The spectrum shown in (a) is measured in the D2 setup while spectra in (b) and (c) are recorded in the D1 setup.
In panels (a) and (b), spectra for larger $t_c$ are scaled by 0.1 for better visualization. The $\gamma$-rays originating from room background are indicated with stars. }
\end{figure}

Figure~\ref{fig1} shows the $\gamma$-ray spectra of the irradiated Torlon 4203, Torlon 4301 and Teflon samples at different times ($t_c$) after the neutron irradiation. Both Torlon 4203 and 4301 samples were exposed to similar neutron dose, while the neutron dose received by the Teflon sample was $\sim40\% $ lower. It can be seen that the most dominant gamma-ray is 511~keV in all the three samples, but the Torlon and Teflon samples show different levels of activity and different impurities. The Teflon and Torlon samples contain fluorocarbon in different proportions, which is reflected in the intensity of the 511~keV $\gamma$-ray with Teflon having the maximum intensity.

\begin{figure}[H]
\includegraphics[scale=0.6]{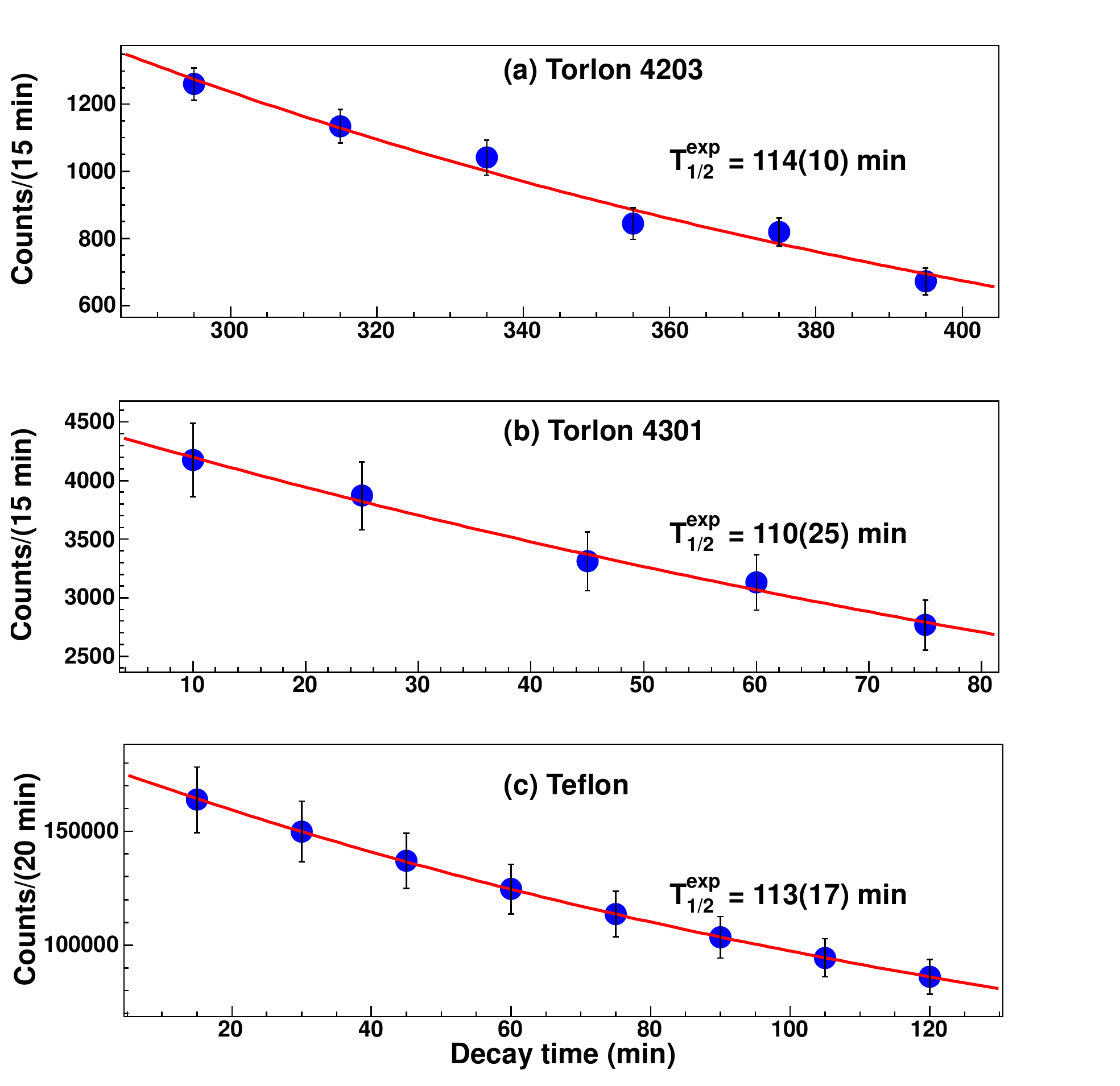} 
\begin{centering}
\caption{\label{fig2}(Color online) Decay curves for 511~keV $\gamma$-ray of $\rm^{18}F$ formed in the neutron irradiated (a) Torlon 4203 with $E_p$ = 20~MeV and $t_{irr}$ = 10~h, (b) Torlon 4301 with $E_p$ = 20~MeV and $t_{irr}$ = 3~h and (c) Teflon with $E_p$ = 20~MeV and $t_{irr}$ = 2~h.}
\end{centering}
\end{figure}

\begin{figure}[H]
\includegraphics[scale=0.5]{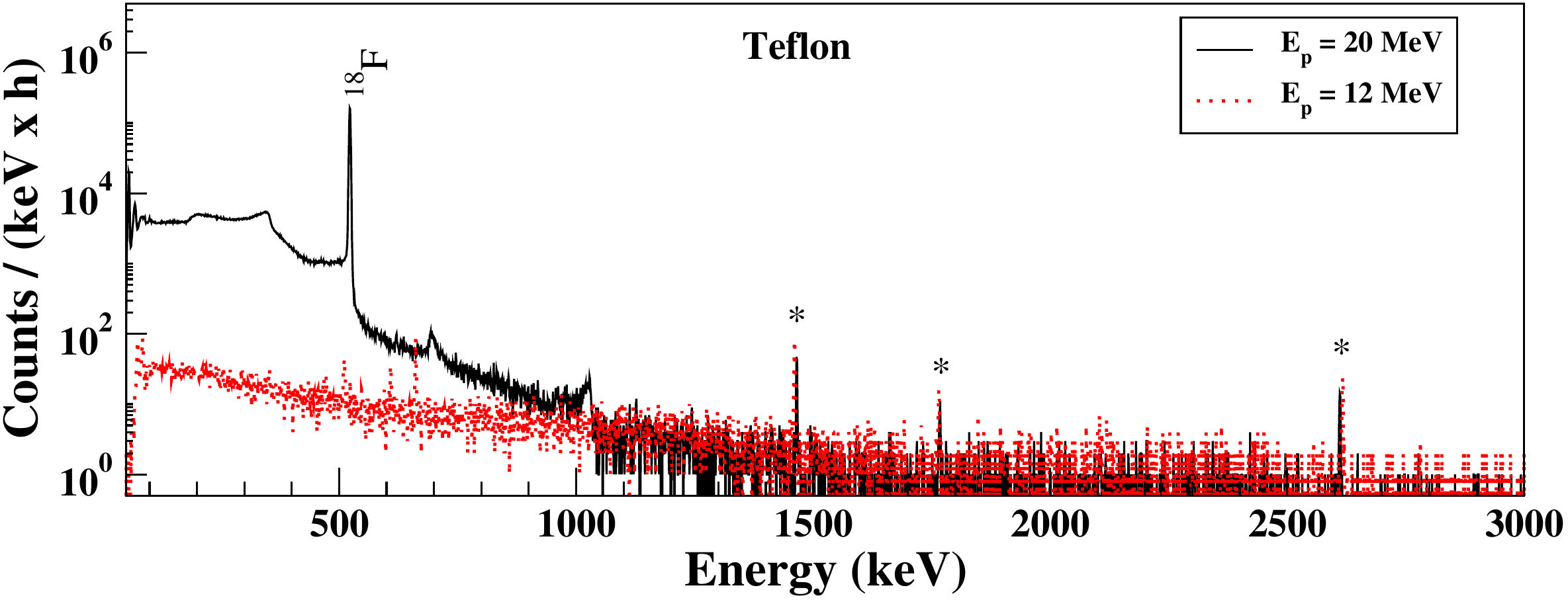} 
\centering
\caption{\label{fig3}(Color online) $\gamma$-ray spectra of the neutron irradiated Teflon with $E_p$ = 20~MeV, $ t_{irr} =$ 2~h and $ N_n\sim 0.82(6)\times10^{10}$ (shown by solid black lines) together with $E_p$ = 12~MeV, $ t_{irr} =$ 13~h and $ N_n\sim 1.27(9)\times10^{10}$ (shown by dotted red lines). Both the spectra have been measured after similar cooling time ($t _{c} $) 14~min and 10~min, respectively, in the D1 setup. Stars have same meaning as in figure 1.}
\end{figure}

\begin{figure}[H]
\includegraphics[scale=0.5]{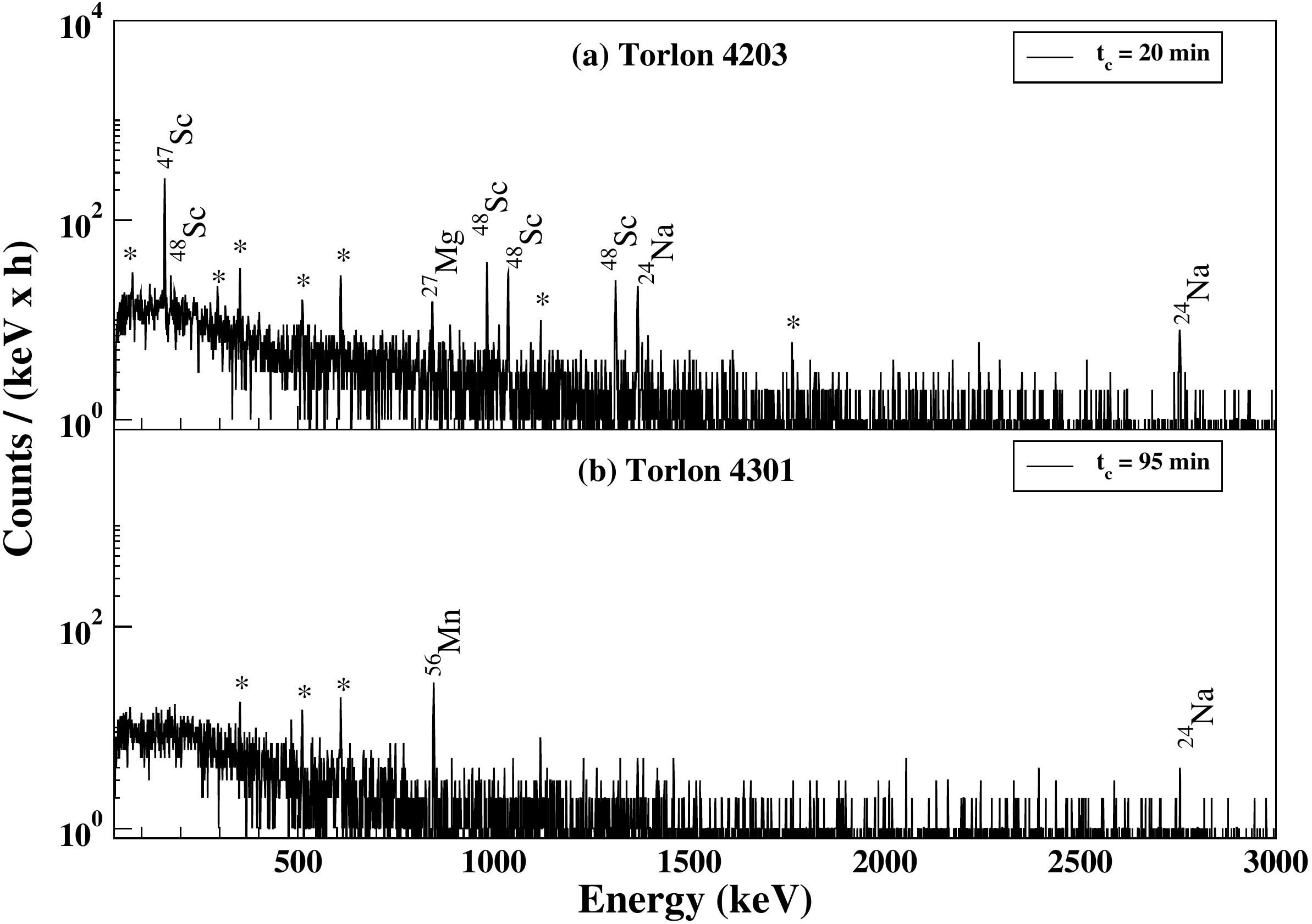} 
\centering
\caption{\label{fig4}(Color online) $\gamma$-ray spectra of the neutron irradiated (a) Torlon 4203 ($ N_n\sim 1.4(1)\times10^{10}$) and (b) Torlon 4301 ($ N_n\sim 1.3(1)\times10^{10}$) with $E_p$ = 12~MeV and $ t_{irr}= $ 13~h. Both the spectra are recorded in the low background setup. Stars have same meaning as in figure 1.}
\end{figure}

As mentioned earlier, the Torlon 4203 contains TiO$_2$ \cite{solvay} and many gamma-rays originating from Ti(n, X)Sc reactions are clearly visible (see table~\ref{listBe}).  
Most of the Sc isotopes formed are short-lived and produce $\gamma$-rays with $E_{\gamma}<1312$ keV. However, $^{46}$Sc has a relatively long half-life, namely, $T_{1/2}$ = 83.79~d.  
It may be mentioned that the Large Underground Xenon (LUX) dark matter experiment has observed background from $^{46}$Sc, which was formed due to the cosmogenic activation of the LUX Titanium cryostat~\cite{luxbkg}. 
In case of the Torlon 4301, $\gamma$-rays resulting from $\rm^{56}Fe(n,p)\rm^{56}Mn$ reaction were observed (see figure~\ref{fig1}(b)). 
Both the Torlon samples have traces of Al, which gives rise to $\gamma$-ray of energy 2754.0~keV (which is higher than $Q_{\beta\beta}(^{124}Sn) $) with a $T_{1/2}$ = 14.99~h and is highly undesirable. Figure~\ref{fig2} shows the decay curves for 511~keV $\gamma$-ray in the irradiated Torlon and Teflon samples. The background rate at 511~keV in the different detector systems has been taken into account. The origin of 511~keV as the $\beta^+$ annihilation $\gamma$-ray from the $\rm^{19}F(n,2n)\rm^{18}F$ reaction is confirmed since the measured half-life agrees with that of $^{18}$F within errors, namely, $T_{1/2}^{ref}$ = 109.77(5)~min~\cite{nndc}.
Considering the threshold energy $E_n\sim$ 11.5~MeV for $\rm^{19}F(n,2n)\rm^{18}F$ reaction~\cite{nndc}, this channel is not expected to be activated at lower neutron energies. The $\gamma$-ray spectra of irradiated samples at $ E_{p} =$ 12 MeV are shown in figure~\ref{fig3} for Teflon (in dotted red lines) and in figure~\ref{fig4} for the Torlon samples. It can be clearly seen that the $\rm^{18}F$ is not populated at $E_n \le$ 9.9~MeV ($E_p$ = 12~MeV) and yield of 511 keV $\gamma$-ray is significantly reduced, whereas most of the reaction channels in Torlon are populated even at lower neutron energy (see table~\ref{listBe}). It may be mentioned that the observed peaks at 1022~keV and $\sim$685.6~keV in the Teflon spectrum, originate from summing of two 511~keV $\gamma $-rays and from summing of 511~keV with backscattered gamma-rays, respectively. This is also seen in the $\rm^{nat}Cu$ sample.

Even though the 511~keV $\gamma$-ray activity in Teflon is significantly larger ($\sim$15.5 (1.2) times) than that in Torlon 4301, there is no gamma background at energies higher than 511~keV  in Teflon. Therefore from the neutron-induced gamma background consideration, Teflon seems to be a better candidate as compared to the Torlon for use in the TIN.TIN detector. 
 
\subsection{Neutron-induced activity from $\rm ^{nat}Pb $ and $\rm ^{nat}Cu $}

\begin{figure}[H]
\includegraphics[scale=0.6]{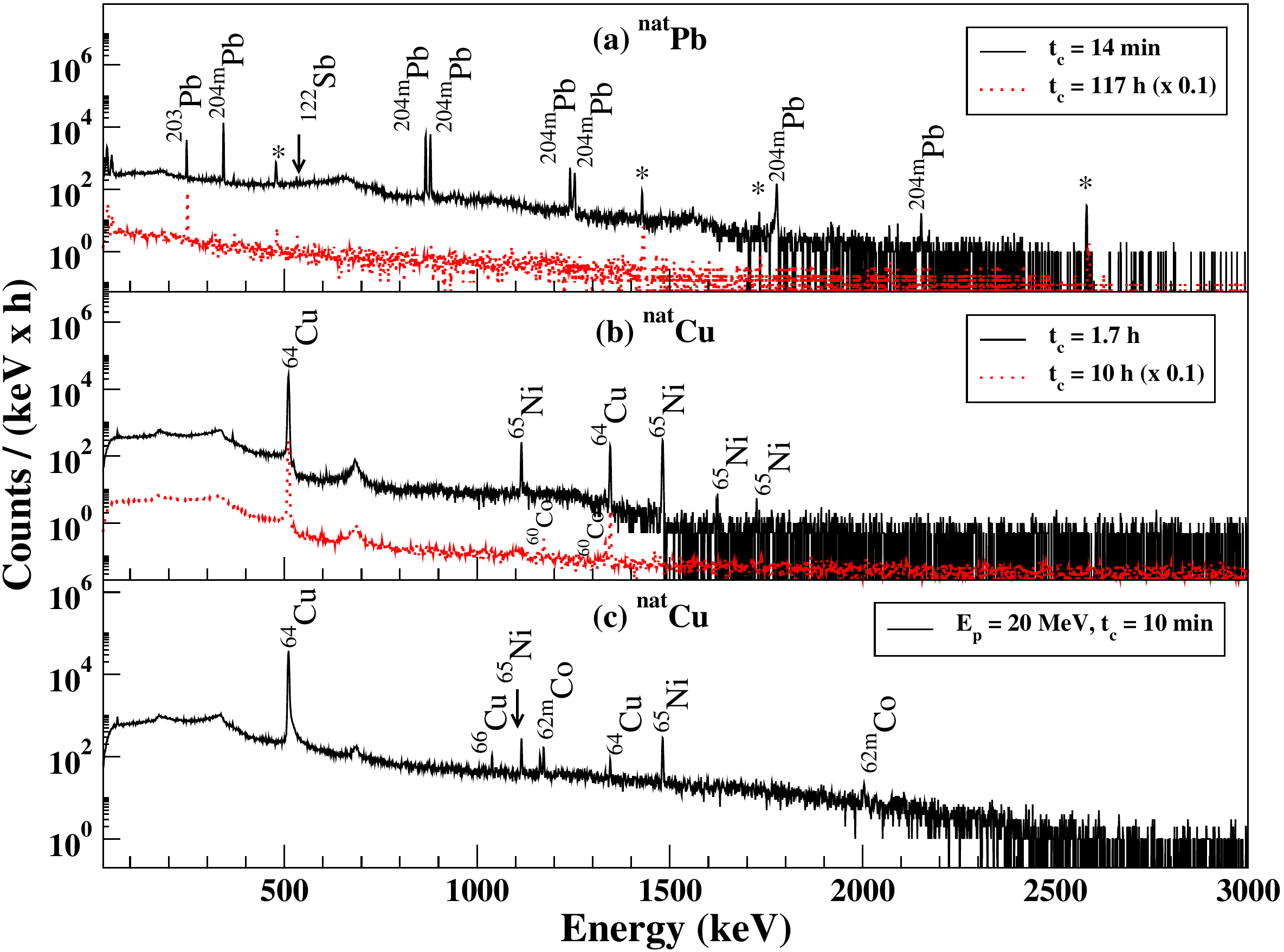} 
\begin{centering}
\caption{\label{fig5}(Color online) $\gamma$-ray spectra of the neutron irradiated (a) $\rm^{nat}Pb$ with $E_p$ = 20~MeV, $t_{irr}$ = 2~h and $  N_n\sim 0.69(5)\times10^{10}$ for different cooling time $t_c$, (b) $\rm^{nat}Cu$ with $E_p$ = 20~MeV, $t_{irr}$ = 10~h, $ N_n\sim 5.9(4)\times10^{10}$ for different cooling time $t_c$ and (c) $\rm^{nat}Cu$ with $E_p$ = 20~MeV, $t_{irr}$ = 2~h and $ N_n\sim 1.09(8)\times10^{10}$. The spectrum shown in (a) is measured in the D2 setup while those in (b) and (c) in the low background setup. In panels (a) and (b), spectra for larger $t_c$ are scaled by 0.1 for better visualization. Stars have same meaning as in figure 1.}
\end{centering}
\end{figure}

The Lead shield is generally closer to the detector assembly and the gamma-rays produced by neutron-induced reactions in Lead can contribute to the background levels. It has been previously reported in Ref.~\cite{lead} that inelastic scattering of neutrons in Lead can be a significant source of background for double beta decay experiments.
The $\gamma$-ray spectra of the irradiated~$\rm ^{nat}Pb $ and $\rm ^{nat}Cu $ samples are shown in figures~\ref{fig5} (a) and (b), respectively. 
 Gamma-rays originating from decay of $\rm^{203}$Pb and $\rm^{204m}$Pb (see table~\ref{listBe}) are seen in the spectrum. In addition, Sb impurities are also found in the Lead sample. It should be noted that the decay of $\rm^{124}$Sb produces many gamma-rays with energies higher than $ Q_{\beta\beta} $ of $\rm^{124}$Sn but with small branching fractions: 2294.0~keV (0.0320$ \% $), 2323.5~keV (0.00243$ \% $), 2455.2~keV (0.0015 $ \% $), 2681.9~keV (0.00165$ \% $), 2693.6~keV (0.0030 $ \% $) and 2807.5 keV (0.00147$ \% $)~\cite{nndc}. In the present work, only 602.7~keV is observed in the gamma-ray spectrum above the detection limit of the low background setup. But $E_\gamma$ = 2294.0~keV may be a crucial source of background in an underground laboratory with improved sensitivity. 

In the $\gamma$-ray spectrum of $\rm ^{nat}Cu $ (see figure~\ref{fig5}(b)) short-lived activities ($T_{1/2} \sim$ h) such as $\rm ^{64}Cu $ and $\rm ^{65}Ni $ are seen. The long-lived products like $\rm^{60}$Co ($T_{1/2}$ = 5.27~y) are visible in the spectra after sufficient cooling time $ \sim $10~h, when the overall gamma background level due to the decay of the short-lived nuclei is reduced. Short-lived products ($T_{1/2} \sim $ min) such as $\rm^{62m}$Co and $\rm^{66}$Cu formed in the Copper sample are visible where spectra could be measured after shorter cooling time.
 
\begin{figure}[H]
\includegraphics[scale=0.5]{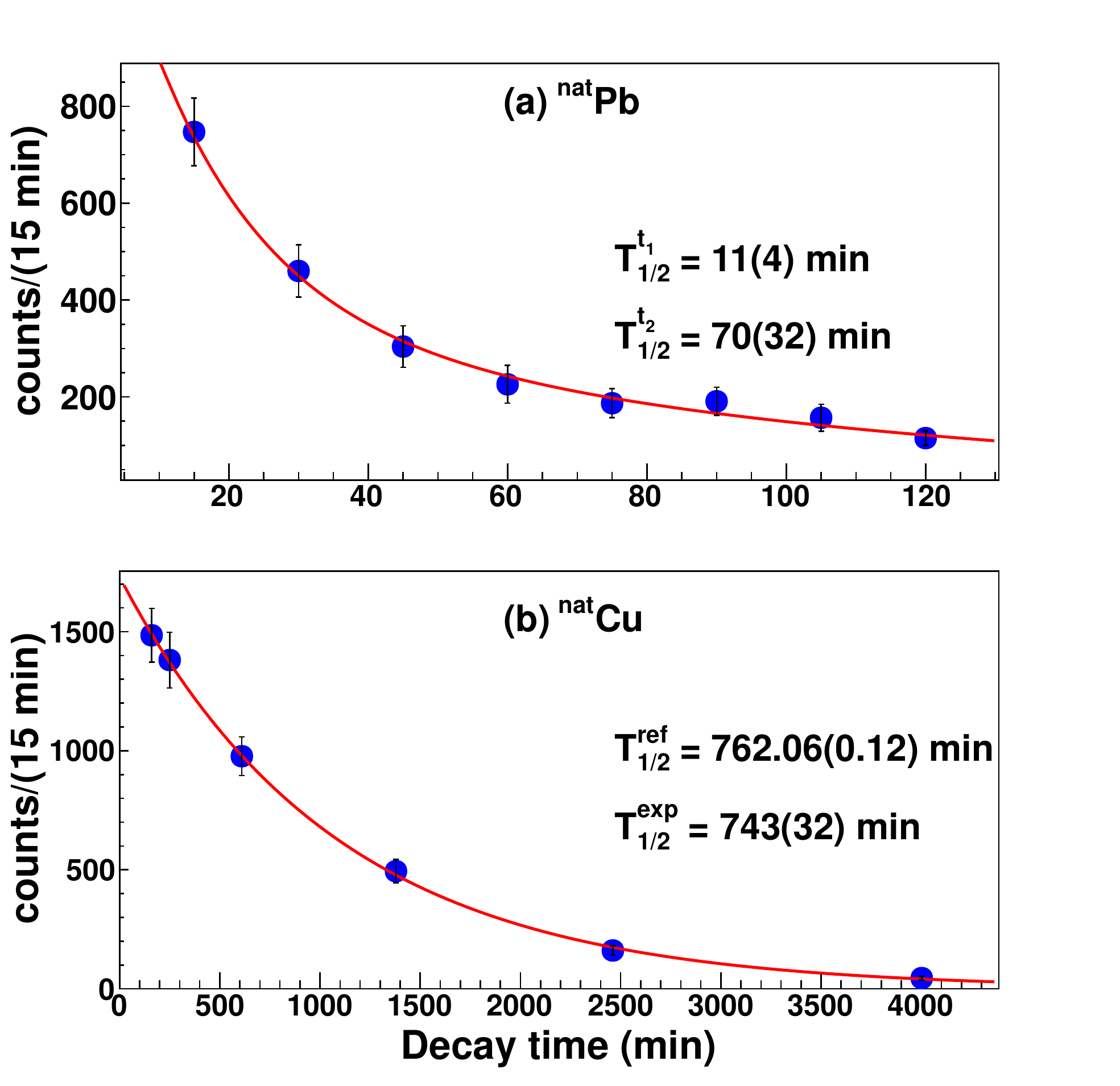} 
\begin{centering}
\caption{\label{fig6}(Color online) Decay curves for 511~keV $\gamma$-ray formed in the neutron irradiated (a) $\rm^{nat}$Pb with $E_p$ = 20~MeV and $t_{irr}$ = 2~h together with two exponential fit, (b) $\rm^{nat}$Cu with $E_p$ = 20~MeV and $t_{irr}$ = 2~h.}
\end{centering}
\end{figure}

Figure~\ref{fig6}(a) shows a decay curve of 511~keV $\gamma$-ray in $\rm^{nat}$Pb sample. A single exponential fit indicated $T_{1/2}\sim$ 41(4)~min, while a two component fit resulted in $ T_{1/2}^{t_1} $ and $ T_{1/2}^{t_2} $ as 11(4) and 70(32) min, respectively but the origin of 511 keV in $ \rm^{nat} $Pb was not identified. Whereas the decay curve in figure~\ref{fig6}(b) for $\rm ^{nat}$Cu gives $T_{1/2}\sim$ 12.4(5)~h, implying that the 511~keV $\gamma$-ray results from the $\rm^{63}Cu(n,\gamma)\rm^{64}Cu$ reaction. It should be noted that no $\rm^{18}$F was observed in the $\rm ^{nat} $Pb or $\rm ^{nat} $Cu samples, confirming that the Teflon sample holder/spacers did not contribute to observed impurities in these samples.

Refs.~\cite{cuoredesign, nemo} have reported the formation of $\rm^{60}$Co in Copper due to cosmogenic activation. 
In addition, the $\rm^{62m}Co$ decay produces several high energy $\gamma$-rays (see figure~\ref{fig5}(c)). Therefore for minimizing the Co activity, it is essential to store Copper in an underground location. 

\subsection{Neutron-induced activity in $\rm ^{nat}Sn $ and $\rm ^{124}Sn $}

\begin{figure}[H]
\includegraphics[scale=0.55]{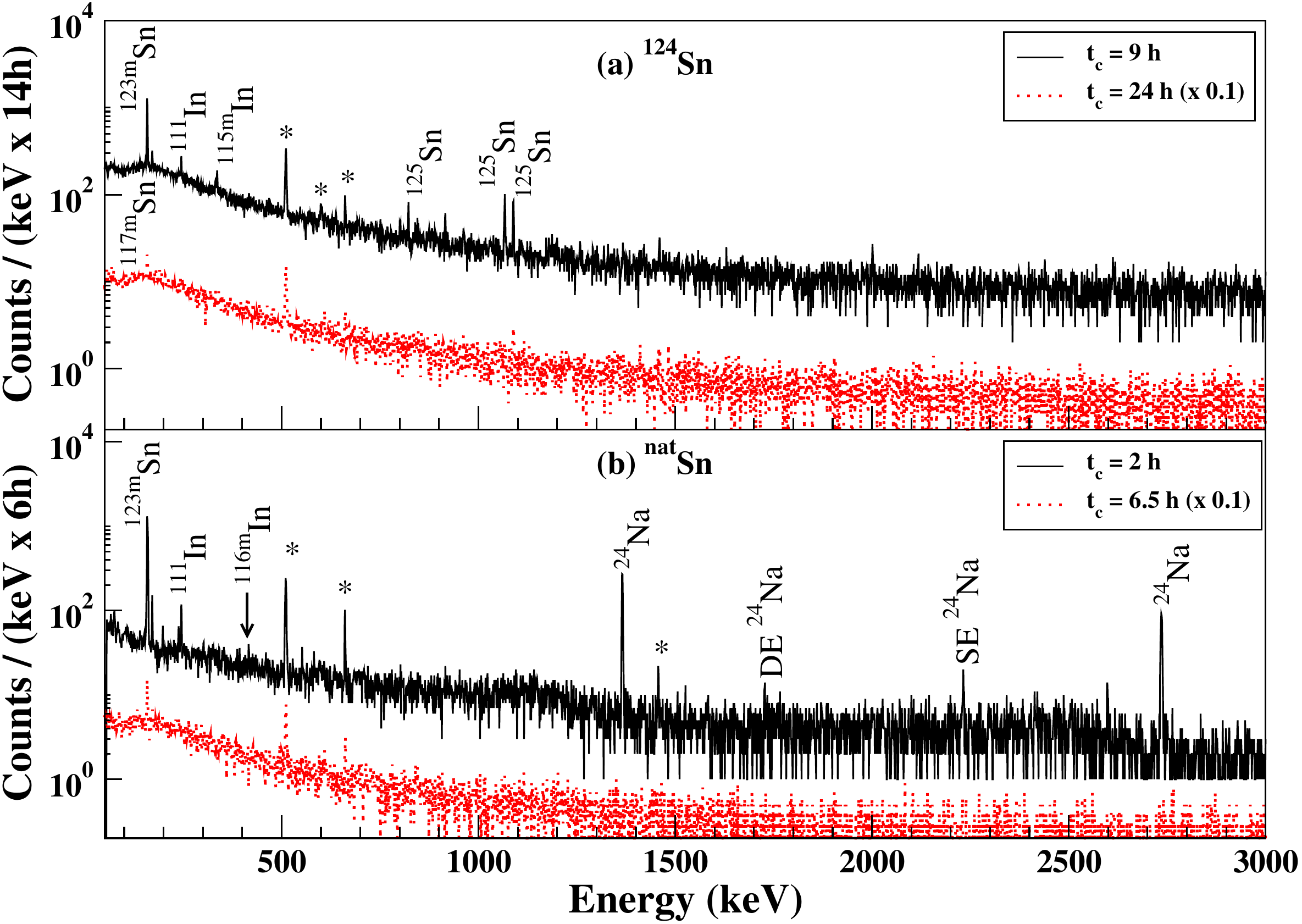} 
\begin{centering}
\caption{\label{fig7}(Color online) $\gamma$-ray spectra of the neutron irradiated (a) $\rm^{124}$Sn with $E_p$ = 20~MeV, $t_{irr}$ = 10~h, $  N_n\sim 4.1(3)\times10^{10}$ (solid black lines) and with $E_p$ = 10 MeV, $t_{irr}$ = 5~h and $ N_n\sim 0.44(4)\times10^{10}$ (dotted red lines), (b) $\rm^{nat}$Sn with $E_p$ = 20~MeV, $t_{irr}$ = 2~h, $  N_n\sim 1.9(1)\times10^{10}$ (solid black lines) and with $E_p$ = 10~MeV, $t_{irr}$ = 5~h, $ N_n\sim 0.51(4)\times10^{10}$ (dotted red lines). All spectra are recorded in the low background setup and  those corresponding to larger $t_c$ have been scaled by 0.1 for better visualization. Stars have same meaning as in figure 1.}
\end{centering}
\end{figure}

Figure~\ref{fig7}(a) shows the $\gamma$-ray spectra of the neutron irradiated ($E_{max}=$17.9~MeV) enriched $\rm^{124}$Sn (97.2$\%$) sample. In addition to the gamma-rays originating from neutron activation of $\rm^{124}$Sn, reaction products of other Sn isotopes, namely, $\rm^{112}$Sn, $\rm^{115}$Sn, $\rm^{116}$Sn, $\rm^{117}$Sn and $\rm^{122}$Sn, are also found in the enriched sample (see  table~\ref{listBe}). Most of the isotopes formed are short-lived, the longest-lived being $\rm^{123}$Sn with a $T_{1/2}$ = 129.2~d. The highest energy gamma-ray $E_\gamma$ = 2112.3~keV originates in the decay of $\rm^{116m}$In. Some of the observed reaction products  can be produced by different Tin isotopes depending on the incident neutron energy and the relative cross-sections. For example,   $\rm^{123m}$Sn can be formed either by $\rm^{122}Sn(n,\gamma)^{123m}Sn$ or by $\rm^{124}Sn(n,2n)\rm^{123m}Sn$ reaction. The contribution from $\rm^{122}$Sn was probed by low energy neutron irradiation ($E _{max} $ = 7.9~MeV corresponding to $E _{p} $ = 10~MeV) where the $\rm^{124}Sn(n,2n)\rm^{123m}Sn $ reaction is unfavoured.
The observation of significantly reduced ($0.16\%$) but measurable activity of $\rm^{123m}$Sn ($E_\gamma$ = 160.3~keV) at lower neutron energy clearly indicated the traces of $\rm^{122}$Sn in the enriched sample.
Similarly, $\rm^{115m}In$ ($E_{\gamma}$ = 336.2 keV) can be produced from $\rm^{115,116}$Sn with high energy neutrons but at lower neutron energy only $\rm^{115}In(n,n')\rm^{115m}In$~~($ \rm^{115}In $ natural abundance 95.7$ \% $) is the possible reaction channel. Thus, observation of 336.2 keV with low energy neutrons implies presence of trace impurity of $\rm^{115}$In in the enriched Tin target. 
In the observed spectra, $\gamma$-rays 1088.6 and 1089.2~keV originating from decay of $\rm ^{123} $Sn ($ T_{1/2} $ = 129.2~d) and $\rm ^{125} $Sn ($ T_{1/2} $ = 9.64~d), respectively, could not be separated.
Measurements after $ t_{c} \sim$~10~d showed that the relative yield of $ E_{\gamma}$ = 1089.2~keV was higher than that for $ E_{\gamma}$ = 1067~keV confirming the formation of $\rm^{123} $Sn. It should be noted that 331.9 keV and 822.5~keV $\gamma$-rays from $\rm ^{125m}$Sn and $\rm^{125}$Sn, respectively, were also visible.

Figure~\ref{fig7}(b) shows the $\gamma$-ray spectra of irradiated $\rm^{nat}Sn$ sample with low energy neutrons, where the gamma-rays from reaction products of $\rm^{112}$Sn, $\rm^{116}$Sn and $\rm^{122}$Sn are visible. The other stable isotopes of Sn upon neutron activation form either long-lived and/or stable reaction products and hence could not be observed. 
It should be noted that 336.2~keV $ \gamma $-ray from $\rm ^{115m} $In was not visible in the $\rm^{nat} $Sn (7N) sample at the same detection sensitivity as in case of $\rm^{124} $Sn. Gamma-rays originating from decay of $^{24}$Na was observed in the samples irradiated in an Al target holder, produced via $\rm^{27}Al(n,\alpha)\rm^{24}Na$ reaction (see figure~\ref{fig7}(b)). No additional impurities are seen in the $\rm^{nat}$Sn (7N) sample.


\subsection{Effect of neutron-induced gamma background for $ 0\nu\beta\beta $ decay in $ \rm^{124} $Sn}
Neutron-induced gamma background at energies $E_{\gamma}\ge2.1\rm~MeV$ is estimated for the measured neutron flux corresponding to $E_{p}=20 $ MeV. Activities of different reaction products in the $\rm^{nat}Cu$, $\rm^{nat}Pb$ and $\rm^{124, nat}Sn$ samples are calculated from the yields of observed $\gamma$-rays of 2003.7~keV, 602.7~keV and 416.9~keV, respectively~(see table~\ref{listBe}). These $\gamma$-rays could be observed only in the close counting geometry even in the low background setup (high efficiency), in case of higher intensities in the respective decay chains. The activity thus obtained for a particular reaction product was then used to estimate the expected background from gamma rays in the ROI using known branching ratios~(see table~\ref{listBe}). Table~\ref{bkg} shows the expected yield of such high energy gamma rays in the $\rm^{nat}Cu$, $\rm^{nat}Pb$ and $\rm^{124, nat}Sn$ samples. The neutron flux is corrected for solid angle subtended by targets in the cascade geometry, placed at different distances ($d$) from the production target. It should be noted that the coincident summing of low energy gamma-rays in these decay cascades can also produce gamma background in the ROI, which will depend on the detector configuration.

\begin{table}[!h]
\centering
\caption{\label{bkg} Estimated neutron-induced background from the high energy $\gamma$-rays in Pb, Cu and Sn samples. }
\begin{tabular}{ |c |c|c| c|c|c|}
\hline
Sample  & Neutron fluence&Reaction&$ T_{1/2} $ & $E_\gamma$ of interest &Expected Intensity of $E_\gamma$ \\ 
& $ n~cm^{-2} (\times 10^{10})$ & Product& &(keV) & ($Bq~g^{-1}) $ \\ \hline

\multirow{2}{*} {$\rm^{nat}Pb$} &\multirow{2}{*} {0.30(2)}&\multirow{2}{*} { $\rm^{124}Sb$ }& \multirow{2}{*} {60.2~d} & 2182.6  & 0.0007(3) \\

 && &   & 2294.0 & 0.0005(2)\\\hline

\multirow{2}{*} {$\rm^{nat}Cu$} & \multirow{2}{*} {0.33(2)}& \multirow{2}{*} {$\rm^{62m}Co$} & \multirow{2}{*} {13.91~min} & 2301.9  & 6(2)\\
 
&&  &&  2882.3  & 4(1) \\\hline

$\rm^{124}Sn$& 1.6(1)  & $\rm^{116m}In$ & 54.29~min &2112.3 & 5(1)\\\hline

$\rm^{nat}Sn$& 0.84(6)&$\rm^{116m}In$& 54.29~min&2112.3& 24(6)  \\ \hline
\end{tabular}
\end{table}
Most of the activities producing high energy gamma-rays are short-lived and can be minimized by storage in an underground location prior to use in the detector setup. Typical neutron flux in underground locations at $E_{n} <10 \rm~MeV $ is $ 10^{-6} n~cm^{-2}s^{-1} $~\cite{bellini} and the required overall background level will be $ < $10$ ^{-2} $~counts/(keV~kg~year). Hence, contribution from Cu and Pb samples in the region of high energy gamma-rays would be negligible. From table~\ref{bkg}, it can be seen that $\rm^{nat}Sn$ will produce $ \sim 5(2)$ times higher gamma background of 2112.3~keV on neutron activation and can be of concern. 

\section{Conclusions}
Neutron-induced background has been studied in various materials to be used in the TIN.TIN detector, which is under development for the search of $ 0\nu\beta\beta $ decay in $\rm^{124}Sn$. In the present work, materials such as Torlon 4203 and 4301, Teflon, $\rm^{nat} $Cu, $ \rm^{nat} $Pb and $\rm^{124, nat}Sn$ are studied. The contribution to the gamma background has been evaluated for an average neutron flux $\sim$ 10$^{6}~n~cm^{-2}s^{-1}$ integrated over neutron energy $ E_{n}$ = $ \sim $ 0.1 to $ \sim $18~MeV. Both Torlon samples show presence of Al which will contribute to high energy gamma background. In addition, Torlon 4301 has Fe impurity while Ti in Torlon 4203 can produce long-lived impurities like $^{46}$Sc. Teflon shows only 511~keV $\gamma$-ray activity resulting from $\rm^{19}F(n,2n)\rm^{18}F$ reaction at $E_n\ge$ 11.5~MeV. Hence, Teflon appears to be a better material for support structures in the Sn cryogenic bolometer from neutron-induced background consideration. Although, the $\rm^{nat}Cu$ sample and Sb impurity in $\rm ^{nat} $Pb produces high energy gamma background ($ E_{\gamma} > 2.1\rm~MeV$) upon neutron activation, the contribution in the ROI of $ 0\nu\beta\beta $ decay in $\rm^{124}$Sn is estimated to be negligible. The neutron-induced reactions form short-lived activities in both $\rm^{124}$Sn and $\rm^{nat}$Sn samples, which are of concern for the Tin detector. Among the various Sn isotopes formed $\rm ^{123} $Sn has the longest half-life $ T_{1/2} $ = 129.2~d, while
$\rm^{116m}$In produces high energy $\gamma$-ray of 2112.3~keV. Thus, for background reduction enriched Tin is preferable as compared to natural Tin. Further, these results suggest that it would be necessary to store Sn material in underground location for extended periods prior to use in the cryogenic bolometer setup.

\acknowledgments
The authors would like to thank Mr.~S.C.~Sharma, Mr.~M.S.~Pose, Mr.~K.V.~Anoop and Mr.~K.V. Divekar for help during the setup and Mr.~A.G.~Mahadkar for the preparation of targets. We are grateful to the Pelletron Linac staff for the smooth operation of the machine during the experiment. 


\end{document}